\documentclass[11pt,onecolumn,amssymb,nofootinbib]{revtex4}
\usepackage{amsmath, amsthm, amscd, amssymb}
\usepackage{graphicx, braket}
\usepackage{bm}
\usepackage{bbm}
\usepackage{epstopdf}

\begin{document}

\title{\bf Asymptotic analysis of the ``simulated horizon'' segment of the Collins spiral } \bigskip

\author{Stephen L. Adler}
\email{adler@ias.edu} \affiliation{Institute for Advanced Study,
1 Einstein Drive, Princeton, NJ 08540, USA.}

\begin{abstract}
The Tolman-Oppenheimer-Volkoff (TOV) equations for a  massless fluid  take the form of a pair of coupled autonomous first order differential equations, which can be employed in a  model for a ``dynamical gravastar'' black hole mimicker.  The mimicker has no true horizon, but rather a ``simulated horizon'', outside which the geometry resembles a Schwarzschild black hole, but inside which the $g_{00}$ component of the metric is always positive and becomes exponentially small.  Collins has reinterpreted the relevant TOV equations in terms of a two-dimensional flow with a spiral form, and Z\"ollner  and K\"ampfer have mapped the simulated horizon to a specific segment of the Collins spiral.  We give here results of an asymptotic analysis, relating initial  values at the small radius end of this spiral segment to the black hole mimicker mass and other parameters that emerge at the large radius kink in the TOV solution, which corresponds to the simulated horizon.  A curious feature of this asymptotic mapping, given in Sec. IIB, is the appearance of power law behaviors with exponents of $1/5$ and $1/10$.
\end{abstract}

\maketitle
\section{Introduction}
\subsection{TOV solution represented as a Collins spiral }

Studies of spherically symmetric compact astrophysical objects, such as white dwarf stars and neutron stars, conventionally start \cite{zeld}, \cite{camen} from the Einstein equations rewritten in the form of the TOV equations \cite{oppen}.  For a  relativistic fluid, with pressure $p$ and  energy density $\rho$  related by a  ``$\gamma$-law'' equation of state $p = (\gamma-1) \rho$ (of which a massless relativistic fluid with $\gamma=4/3~,~~~p=\rho/3$ is a special case), the TOV equations take the form of a pair of planar autonomous equations.  This facilitates their graphical study, and Collins \cite{coll} has plotted the phase space flow lines of an equilibrium relativistic $\gamma$-law  fluid with spherical, plane, or hyperbolic symmetry. The spherical symmetry case, of relevance to compact astrophysical objects, shows characteristic spiral flow lines, hence the term ``Collins spiral''.

\subsection{The  black hole mimicker ``simulated horizon''}

In a series of papers reviewed in \cite{adler1}, Adler has proposed a model for a horizonless black hole mimicker, constructed by solving the TOV equations for assumed outer  and inner  equations of state.  The model  consists of an outer shell of relativistic fluid with $p=\rho/3$, surrounding an inner core generated by a high pressure phase transition, with the core fluid obeying an  equation of state $p+\rho=\epsilon>0$ with $\epsilon$ a small constant. In the limit of zero $\epsilon$, this is the ``vacuum'' equation of state proposed initially by Gliner \cite{gliner}  for a black hole interior, and utilized and popularized by Mazur and Mottola \cite{mazur} in their papers on gravastars (GRAvity VAcuum STARS).  In the Mazur--Mottola model, the Gliner equation of state is attained via a pressure jump at specified radii. The  Adler model, noting that the Einstein and TOV equations require  pressure to be continuous when all physical quantities are bounded, postulates a first order phase transition jump to negative energy density $\rho$ in the internal core.  Such a jump is allowed by the TOV equations, and by renormalized quantum theory.  This results in a black hole mimicker with a ``simulated horizon'', which closely approximates a Schwarzschild solution outside a smooth boundary, but has a metric component $g_{00}$ that remains positive and decays exponentially to zero at and inside the simulated horizon.  In a subsequent paper, Adler and Doherty \cite{adler2} studied the simulated horizon as a property of the outer relativistic fluid layer, rewriting the TOV equations there, expressed in terms of dimensionless variables, as a pair of coupled autonomous equations which define a phase space flow.

\subsection{The mimicker model reinterpreted as a ``core-corona'' model, with  the  ``simulated horizon''   mapped into a segment of the Collins spiral}

In  Appendix B  to their  paper \cite{zollner} developing a ``core-corona'' model for neutron stars, Z\"ollner and K\"ampfer have reinterpreted the Adler black hole mimicker model in the core-corona framework, with the ``corona'' the relativistic fluid outer layer with $p=\rho/3$, and the ``core'' the Gliner-like interior state with $p+\rho \simeq 0$.  Using the TOV equations for the corona in autonomous equation form, they show in their Figure 10, how the mimicker ``simulated horizon'' maps into a segment of the Collins spiral. In the example in this figure, the inner boundary of the relativistic fluid has positive parameters which obey  inequalities given in \cite{adler1}, and so  suffices to give a simulated horizon on a Collins spiral plot with logarithmic coordinates.  The bunching of flow lines seen in Fig. 10 of \cite{zollner} is a reflection of the fact that autonomous system flow lines cannot cross, and results in limiting behaviors used in Sec. IIC below.   We will be interested here in the case in which the inner boundary parameter representing the volume integrated energy density takes very large negative values, with the the aim of giving asymptotic formulas valid in this limit.

\subsection{Relativistic fluid (corona region) TOV equations rewritten as an autonomous system in terms of scale invariant variables}
 The TOV  equations for a spherically symmetric fluid with pressure $p(r)$ and energy density $\rho(r)$,  take the  general form

\begin{align}\label{newTOV}
\frac{dm(r)}{d r}=&4\pi r^2\rho(r)~~~,\cr
\frac{dp(r)}{d r}=&-\frac{\rho(r)+p(r)}{2} \frac{d\nu(r)}{d r} ~~~,\cr
\frac{d\nu(r)}{d r}=&\frac{N(r)}{1-2m(r)/r}~~~,\cr
N(r)=&(2/r^2)\big(m(r)+4\pi r^3 p(r)\big)~~~.\cr
\end{align}
Here  $m(r)$ is the volume integrated energy density within radius $r$,   and $\nu(r)=\log\big(g_{00}(r)\big)$.   The  general form TOV equations become a closed system when  supplemented by an equation of state $\rho(p)$ giving the energy density $\rho$ in terms of the pressure $p$.  For a relativistic fluid with  $p(r)=\rho(r)/3$,
the TOV equations can be recast  in a form manifestly invariant under radial scale transformations, by defining  new  quantities
\begin{align}\label{invs}
t=&\log{r} ~~~,\cr
dt =& dr/r ~~~,\cr
\alpha(t)=& m(r)/r~~~,\cr
\delta(t)=&4 \pi r^2 p(r)~~~.\cr
\end{align}
Scale transformation of $r$ corresponds to shifting the origin of $t$, and as discussed in \cite{adler1},  the remaining three quantities $dt$, $\alpha(t)$, and $\delta(t)$ are invariant under this shift.  In terms of these variables, the TOV equations for $p$ and $m$ take the scale-invariant form
\begin{align}\label{scaleinvTOV}
\frac{d\alpha(t)}{dt}=& 3 \delta(t)-\alpha(t)~~~,\cr
\frac{d\delta(t)}{dt}=& -4 \delta(t) \frac{\delta(t)+2 \alpha(t)-1/2}{1-2\alpha(t)}~~~.\cr
\end{align}
The scale-invariant form of the TOV equation for $\nu$  is
\begin{equation}\label{nueq}
\frac{d\nu(t)}{dt}= 2 [\alpha(t)+\delta(t)]~~~,
\end{equation}
which can be integrated in terms of the solution of Eqs. \eqref{scaleinvTOV}.

These equations are what are called ``autonomous'' differential equations, in that no functions of the independent variable $t$ appear in the coefficients multiplying the dependent variables $\alpha(t)$ and $\delta(t)$.  This autonomous equation system describes a 2-dimensional flow,  a feature that is exploited in the analyses of \cite{coll} and \cite{zollner}.

\subsection{An illustratiion of the solutions}

As a concrete illustration of the use of these equations, we plot in Figs. 1 and 2 the results of integrating Eqs. \eqref{scaleinvTOV} from an inner radius chosen as $t=t_0=0$  to an outer radius $t=40$, with the initial values of $\alpha$ and $\delta$ taken as
\begin{align}\label{initval}
\alpha_0=& -10^{18}        ~~~,\cr
\delta_0=&  10^{-10}         ~~~,\cr
\end{align}
for which  $-\alpha_0>>1>>\delta_0$.
For $t<12$  $\alpha$ is strongly negative, and for $6<t<12 $ $\delta$ is strongly positive, and so to keep the region beyond $t=12$ on scale we have started the plots at $t=12$, even though the integrations start at $t=0$.
We see that $\alpha$ has a sharp kink, with peak value $\alpha \simeq 1/2$ at  $ t_*\simeq 13.2164$, which as discussed in \cite{adler1} is the location of the simulated horizon. At the peak, $m(r)\simeq (1/2) e^{t_*} \simeq  274,650 \equiv M$, giving the dynamical gravastar mass $M$.  Above the peak, $m(r)$ decays as $M/r $, becoming smaller than $10^{-4}$ at $t=25$,  until exponentially larger values of the radius $r$ are reached above $t \sim 27.7$.  Above this $t$, which corresponds to a radius $r = e^{27.7} $, more than a million times larger than the simulated horizon radius $r=e^{13.2164}$, $\alpha$ and $\delta$ climb to fixed point values.  Above $t=40$ they are effectively constants, with  $\alpha = 3/14 \simeq 0.214   $ and $\delta = 1/14 \simeq 0.0714$, for which Eq. \eqref{scaleinvTOV} gives  $d\alpha/dt =0~,~~d\delta/dt=0$.

\subsection{Combining the linear plots of Fig. 1 and Fig. 2 into a Collins spiral}

In Fig. 3 we combine the linear plots of Figs. 1 and 2 into a two dimensional parametric plot, with $\alpha$ running along the horizontal axis and $\delta$ running along the vertical axis.  This gives the Collins spiral representation of the solution.  The starting point $t=12$ lies on the short vertical line segment on the  right of Fig. 3.  The  cusp peak of Fig. 1  at $\alpha \simeq 1/2, \,\delta \simeq 1/6, \, t\simeq 13.2164$ lies at the tip of this vertical line segment, and as $t$ increases above $13.2164$ the curve runs down this vertical segment and then along the horizontal axis, continuing to the left hand corner, corresponding to $t\sim 27.7$.  From here the $\alpha$ and $\delta$ values increase towards the fixed point at $\alpha = 3/14, \, \delta = 1/14$, where the solution remains for $t \geq 35$.  The segment of the curve lying along the axes up to the left hand corner corresponds to the simulated horizon and its Schwarzschild-like black hole mimicker exterior.  The interior spiral segment approaching the fixed point corresponds to the region at exponentially large radius $r$, where an isolated  mimicker no longer describes realistic physics.

\section{Asymptotic Analysis}

 \subsection{Left hand and kink peak parameters}

 Starting from any value of $t$ (except the fixed point where evolution stops) the solution to Eqs. \eqref{scaleinvTOV} is uniquely specified by giving the values of $\alpha(t)$ and $\delta(t)$. Taking the starting point of integration on the left as $t_0$, one can give $\alpha_0=\alpha(t_0)$ and $\delta_0=\delta(t_0)$.  Alternatively, at the peak $t_*$ of the kink where $0=d\alpha/dt=3 \delta-\alpha$, the solution can be specified by giving the $t_*$ value where $3\delta(t_*)=\alpha(t_*)$, and the value of $\epsilon$ defined by  $\alpha(t_*)=1/2-\epsilon$.  We turn now to the problem of finding the mapping between these two ways of specifying the solution.

\subsection{Mapping between parameter sets}

In general there is no simple analytic formula for this mapping; one  has to integrate the differential equations numerically. But we will show that in the asymptotic limit of large negative $\alpha_0$ and small  positive $\delta_0$,
\begin{equation}\label{ineq}
-\alpha_0>>1>>\delta_0>0~~~,
\end{equation}
there are useful explicit scaling formulas.  Introducing exponents $x,\,y$ to parameterize the initial values by
\begin{equation}\label{param}
\alpha_0=-10^{\,3+x}~~~,~~~\delta_0=10^{-y}~~~,
\end{equation}
we find the following asymptotic scaling relations, good for $x \geq 5$ and $y\geq 0$  (we have surveyed $5\leq x \leq 15$ and $0 \leq y \leq 10$)
\begin{align}\label{scaling}
t_* \simeq & 1.704 + 2.303 (x+y)/5 ~~~,\cr
M \simeq & \frac{1}{2}e^{t_*} \simeq 2.747 \cdot 10^{(x+y)/5} \simeq 0.6899 (-\alpha_0/\delta_0)^{1/5}~~~,\cr
\epsilon \simeq & 0.007211 \times 10^{-(4x-y)/10} \simeq 0.1143 (\alpha_0^4\delta_0)^{-1/10}~~~.\cr
\end{align}

\subsection{Heuristic derivation of the scaling relations}

We give a heuristic derivation  of the scaling relations, which is not a proof, but we conjecture can be extended into a proof.\footnote{Our guess is that a proof will involve a double limit, taking $\alpha_0 \to -\infty$ and  taking $t_L \to t_0$.}   We start with the $t_*$ and $M$ scaling formulas of Eq. \eqref{scaling}.  When the initial parameters  obey the inequalities of Eq. \eqref{ineq}, the differential equations of Eq. \eqref{scaleinvTOV} simplify  to read
\begin{align}\label{scaleinvTOV1}
\frac{d\alpha(t)}{dt}\simeq & -\alpha(t)~~~,\cr
\frac{d\delta(t)}{dt}\simeq & 4 \delta(t) ~~~,\cr
\end{align}
with solutions, taking $t_0=0$,
\begin{align}\label{scaleinvTOV2}
\alpha(t)\simeq &\alpha_0 e^{-t}~~~,\cr
\delta(t)\simeq & \delta_0 e^{4t}~~~.\cr
\end{align}
Hence  for $t$ not too near the value $t_*$ characterizing the kink in the solution,  say for  $0<t<t_L$ with $t_L$ not too close to $t_*$, the ratio
$R(t) \equiv \alpha(t)/\delta(t)$ obeys
\begin{equation}\label{ratio}
R(t) \simeq (\alpha_0/\delta_0) e^{-5t}~~~.
\end{equation}
In particular, with $t$ at the upper limit $t_L$ of the range where Eq. \eqref{ratio} holds, $R(t_L)$ obeys
\begin{equation}\label{ratio1}
R(t_L) \simeq (\alpha_0/\delta_0) e^{-5t_L}~~~.
\end{equation}
On the other hand, at the coordinate $t_*$ where $0=d\alpha(t)/dt|_{t_*}=3 \delta(t_*)-\alpha(t_*)$, we have
\begin{equation}\label{ratio2}
R(t_*) =3~~~.
\end{equation}
Defining $K(t_L,t_*)$ by
\begin{equation}\label{kdef}
K(t_L,t_*)\equiv R(t_L)/R(t_*)=(\alpha_0/3 \delta_0)e^{-5(t_L-t_*)}e^{-5t_*}~~~,
\end{equation}
after rearrangement we get a formula for $e^{t_*}$,
\begin{equation}\label{final}
e^{t_*}=(-\alpha_0/\delta_0)^{1/5}/[-3 K(t_L,t_*)^{1/5}e^{t_L-t_*}]~~~.
\end{equation}
Rewriting the third line in Eq. \eqref{invs} as
\begin{equation}\label{Meq}
M\equiv m(t_*)= \alpha(t_*) e^{t_*}~~~,
\end{equation}
and noting that near its maximum $\alpha(t)$ takes the value $\alpha(t_*) \simeq 1/2$,
we arrive at a formula for $M$,
\begin{equation}\label{Meq1}
M\simeq \frac{1}{2} e^{t_*} \simeq  (-\alpha_0/\delta_0)^{1/5}/[-6 K(t_L,t_*)^{1/5}e^{t_L-t_*}]~~~.
\end{equation}

Although $K(t_L,t_*)^{1/5}$ is sensitive to the choice of $t_L$, this sensitivity is largely canceled by the factor $e^{t_L-t_*}$, and so  the product $K(t_L,t_*)^{1/5}e^{t_L-t_*}$ is slowly varying with respect to $t_L$.  For a range of values $5\leq x\leq 15$ and $0\leq y\leq 10$,  integrating Eqs. \eqref{scaleinvTOV} from $t=0$ through the kink to determine $t_*$, we find that a suitable choice of $t_L$ is  $t_L \simeq t_*-1$.  From this we get the formulas for $t_*$ and  $M$ in Eq. \eqref{scaling}, with the numerical coefficients accurate to a few tenths of a percent.

We next turn to the scaling formula of Eq. \eqref{scaling} for $\epsilon=\frac{1}{2}-\alpha(t_*)$.   Defining $Q(t)\equiv \alpha(t)^4 \delta(t)$, we learn from Eq. \eqref{scaleinvTOV2} that $Q(t)$ is almost constant from $t=0$ up to a limit $t_L$ just below the kink,  and so
\begin{equation}\label{Qcons}
Q(t_0)\simeq Q(t_L) = Q(t_*) \big(Q(t_L)/Q(t_*)\big)~~~.
\end{equation}
Substituting the evaluations of $Q(t)$ at $t_0=0$ and at $t_*$, and defining $R(t_L,t_*)\equiv  \big(Q(t_L)/Q(t_*)\big)$ ,  this becomes
\begin{align}\label{Rformula}
\alpha_0^4\delta_0=&\frac{1}{3}\big(\frac{1}{2}-\epsilon\big)^5 R(t_L,t_*)\simeq  R(t_L,t_*)/(3 \times 2^5) ~~~,\cr
\end{align}
with $R(t_L,t_*)$ describing the evolution of $Q(t)$ through the kink.  Numerical study of $Q(t)$ indicates that a suitable choice of $t_L$, consistent with obtaining tenths of a percent accuracy, is $t_L =t_*-1.5$.  Recalling that $t_*$ is implicitly fixed as the point where $\delta(t_*)=\alpha(t_*)/3 = (1/2 -\epsilon)/3$, we can compute $R(t_*-1.5,t_*)$ as a function of $\epsilon$ by integrating back from $t_*$ to $t_*-1.5$. By the properties of autonomous systems, this is independent of the  numerical value of the integration origin point $t_*$, which we can conveniently take as $0$.   For $\epsilon$ ranging from $10^{-4}$ to $10^{-10}$, we find a limiting behavior
\begin{equation}\label{limiting}
1/\epsilon\simeq R(-1.5,0)^{1/10}/0.180396~~~.
\end{equation}
Combining this formula with Eq. \eqref{Rformula} yields the $\epsilon$ scaling formula given in Eq. \eqref{scaling}.

\subsection{Further regularities}

We close this section with two further regularities.  The first regularity concerns the magnitude of $\nu(t_0)=\log\big(g_{00}(t_0)\big)$ at the inner boundary with coordinate $t_0$.  From Eq. \eqref{nueq}  we get a formula for $\nu(t_0)-\nu(t_*)$,
\begin{equation}\label{nueq1}\,
\nu(t_0)-\nu(t_*)=\int_{t_0}^{t_*} du 2 [ \alpha(u)+\delta(u)] ~~~.
\end{equation}
Assuming $\delta(u)<<\alpha(u)$, taking $t_*$ to be the effectively infinite radius where $\nu=0$,  using the approximation $\alpha(u) \simeq \alpha_0 e^{-u}$, and eliminating $\alpha_0$ in terms of $M$ by using the second line of Eq. \eqref{scaling}, we get
\begin{equation}\label{nueq2}
\nu(t_0)\simeq 2 \alpha_0 \simeq -2 \delta_0 (M/0.6899)^5 \simeq - 12.80 \delta_0 M^5~~~.
\end{equation}
For the metric component $g_{00}$ at the inner boundary we get
\begin{equation}\label{g00eq}
g_{00}(t_0) \simeq e^{-12.80 \delta_0 M^5}~~~.
\end{equation}
Hence, as asserted above, inside the simulated horizon the metric component $g_{00}$ remains positive, but becomes exponentially small.

The second regularity concerns the width of the interval above the simulated horizon $t_*$ (which lies at $ t_* \simeq 13.2$ in Figs. 1 and 2), where $g_{00}$ decays rapidly towards zero, extending up to the coordinate $t_F$ (with $t_F  \sim 27.7$ in Figs. 1 and 2) where $\alpha$ and $\delta$ climb to their fixed point values.  We have not found a quantitative scaling formula for $t_F-t_*$, but as shown in Table I we observe qualitatively  that as $M$ increases, $t_F-t_*$ also increases, and so for very large $M$ the fixed point is effectively at infinite radius.

\section{Physical application}

Our focus so far has been on mathematical properties of the TOV equations for a relativistic fluid.  We now turn to remarks, necessarily more speculative, on possible applications to astrophysical black holes interpreted as black hole mimickers.  We first note that the variables $\alpha =m(r)/r$ and $\delta=4 \pi             r^2 p(r)$ are of a different character. The scale-free pressure   $\delta$  is intensive in character, and is locally defined at radius $r$.   However, the scale-free mass $\alpha$  is extensive in character, and is cumulative over the entire interior of the mimicker.    So we expect $\delta$ to be limited in range, consistent with how we have parameterized it with  $y$, whereas $\alpha$ can effectively range up to infinity, consistent with how we have parameterized it with $x$.

We next remark on  the natural length scale of our model.  Since we are taking the pressure  pjump inducing a phase transition  as near-Planckian in scale, the natural scale for quantities in our model when dimensions are reinserted is given by  the Planck length, Planck mass, Planck pressure, etc.  Since one solar mass is equal to $\simeq 1.3 \times 10^{38}$ Planck masses, astrophysical black holes, ranging from a few solar masses to $10^{11}$ solar masses, span a range of  $\sim 10^{38}$ to   $\sim 10^{49}$ in Planckian  units.  Hence in Planckian units our model must accommodate exponentially large mass values, which is why we have focused on the asymptotic regime of large scale-free masses in our analysis.  Interpreted this way, our formulas indicate the values $x>>1$  needed for our model to produce mimicker masses corresponding to the range of observed astrophysical black hole masses.

We expect the dimensionless pressure $\delta$ at the jump radius in our model to be larger than its value at the center of a neutron star, since a neutron star cannot have undergone the postulated high pressure phase transition.  In Planckian units the pressure at the center of a neutron star is  $\sim 0.3 \times 10^{-79}$, and a neutron star radius of ten kilometers is $0.6 \times 10^{39}$ in Planck radius units. So the value of $\delta$ at the center of a neutron star is around   $10^{-1}$, corresponding to  $y = 1$.  Hence we expect to  encounter  values  $y \leq 1$ in physical application of our model to black hole mimickers, but not $y>>1$.

\section{Acknowledgement}

I wish to thank  R. Z\"ollner and B. K\"ampfer for an email bringing their work \cite{zollner} to my attention.  I also wish to acknowledge helpful conversations with Tom Spencer,  Mike Weinstein, and
Sarah Brett-Smith.

\begin{table} [b]
\caption{ Trend of $t_F$ (the lower edge of the fixed point region) versus $x$ and $y$, showing that $t_F$ increases with increasing $x,\,y$ and hence with increasing $M$}
\centering
\begin{tabular}{c  c c  }
\hline\hline
 ~~$x$~~ &~~$y$~~&~~$t_F$~~ \\
\hline
5&10&14.6\\
10&10&20.9\\
15&10&27.7\\
\hline
10&5&19.8\\
10&10&20.9\\
10&15&22.1\\
\hline\hline
\end{tabular}
\label{table1}
\end{table}

\vfill\eject

\vfill\eject

\begin{figure}[]
\begin{centering}
\includegraphics[natwidth=\textwidth,natheight=300,scale=1.0]{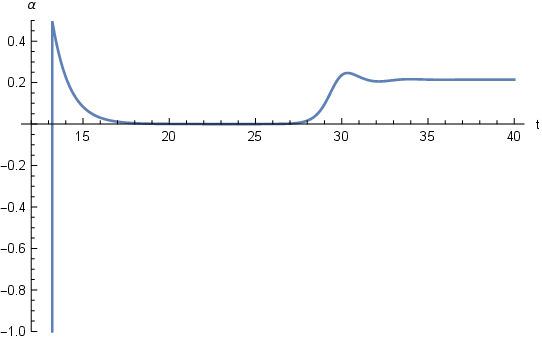}
\caption{Plot of $\alpha$, from just below the kink at $t\simeq 13.2164$ to exponentially large radius values.   }
\end{centering}
\end{figure}

\begin{figure}[]
\begin{centering}
\includegraphics[natwidth=\textwidth,natheight=300,scale=1.0]{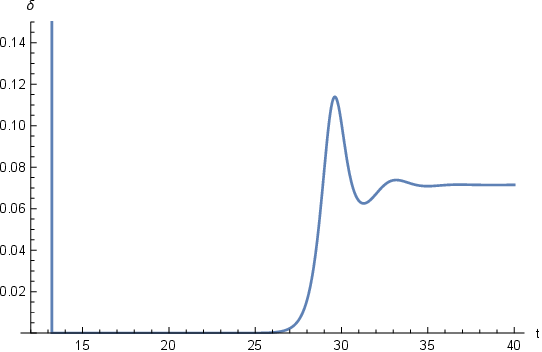}
\caption{Plot of $\delta$, from just below the kink at $t\simeq 13.2164$ to exponentially large radius values.    }
\end{centering}
\end{figure}

\begin{figure}[]
\begin{centering}
\includegraphics[natwidth=\textwidth,natheight=300,scale=1.0]{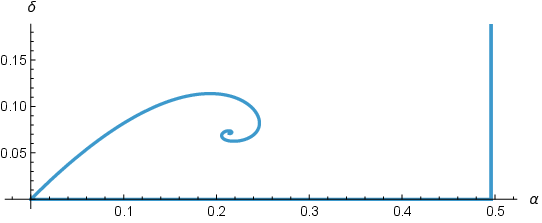}
\caption{Plot of the Collins spiral, obtained by combining the plots of Fig. 1 and Fig. 2 as a planar parametric plot.   }
\end{centering}
\end{figure}


\begin{thebibliography}{99}
\bibitem{zeld} Ya. B. Zeldovich and I. D. Novikov, {\it Stars and Relativity}, The University of Chicago Press (1971), pp. 256-257.
\bibitem{camen}  M. Camenzind, {\it Compact Objects in Astrophysics}, Springer (2007), Secs. 4.1-4.2.
\bibitem{oppen} J. R. Oppenheimer and G. M. Volkoff, {\it Phys. Rev.} {\bf 55}, 374 (1938).
\bibitem{coll}  C. B. Collins, J. Math. Phys. {\bf26}, 2268 (1985).
\bibitem{adler1} S. L. Adler, Mod. Phys. Lett. A{\bf 40}, 2530009 (2025), arXiv:2504.18690.
\bibitem{gliner} E. B. Gliner, {\it J. Exptl. Theoret. Phys.} {\bf 49}, 542 (1965); translation in {\it Sov. Phys. JETP} {\bf 22}, 378 (1966).
\bibitem{mazur}  P. O. Mazur and E. Mottola, ``Gravitational Condensate Stars'', arXiv:gr-qc/0109035 (2001). See also Proc. Nat. Acad. Sci. {\bf 101}, 9545 (2004), arXiv:gr-qc/0407075.
\bibitem{adler2} S. L. Adler and B. Doherty, Int. J. Mod. Phys. D {\bf 34}, 2550056 (2025), arXiv:2309.13380.
\bibitem{zollner} R. Z\"ollner and B. K\'ampfer,  Astronomy {\bf 4}, 10 (2025), arXiv:2506.10032.

\end{thebibliography}
\end{document}